# Universal convolution from wave dynamics: photonic processing and encryption in synthetic dimension

Xiaolong Su,[1] Weiwei Liu,[1,*] Ruiqian Cheng,[1] Haoru Zhang,[1] Xinyao Guo,[1] He Huang,[1] Chengzhi Qin,[1] Peixiang Lu,[1,2,*] and Bing Wang[1,*]

[1]Wuhan National Laboratory for Optoelectronics and School of Physics, Huazhong University of Science and Technology, Wuhan 430074, China

[2]Hubei Key Laboratory of Optical Information and Pattern Recognition, Wuhan Institute of Technology, Wuhan 430205, China

[*]Corresponding authors: W. L. (email: lwhust@hust.edu.cn), P. L. (email: lupeixiang@hust.edu.cn), B. W. (email: wangbing@hust.edu.cn)

## Abstract

Convolution, a cornerstone of signal processing and optical neural networks, has traditionally been implemented by mapping mathematical operations onto complex hardware. Here, we overcome this challenge by revealing that wave dynamics in translation-symmetric lattices intrinsically performs convolution, with the dispersion relation uniquely defining the complex-valued kernel. Leveraging this universal principle, we develop a convolutional architecture of minimal complexity through wave evolution in programmable photonic synthetic lattices, delivering high-throughput, multifunctional capabilities at a rate of 13.5 tera-operations per second (TOPS) for image processing. Beyond convolution acceleration, the kernel's complex nature facilitates the photonic simulation of both irreversible diffusion and reversible unitary quantum dynamics under classical incoherent excitation. Capitalizing on the physics-based reversibility and undetectable phase information, we demonstrate a novel convolution-driven optical encryption strategy. This work establishes a unified framework for photonic computing by grounding convolution in wave dynamics, opening avenues toward scalable, multifunctional photonic processors with high integration potential.

# Introduction

The growing computational demands of artificial intelligence have motivated photonic convolution as a route to surpass electronic limits, achieving remarkable speeds in the tera-operations-per-second (TOPS) regime[1-3]. Various photonic systems, including time-delay networks[4], microring resonators[5-7], and Mach-Zehnder interferometer (MZI) meshes[8,9], have demonstrated considerable success in visual processing tasks, with recent complex-valued schemes further expanding optical computing capabilities[10,11]. However, these implementations generally require complex components to map mathematical operations onto photonic hardware, fundamentally hindering the scalability and integration potential. Moreover, most conventional approaches manipulate photons in time domain, where the processing speed remains ultimately constrained by serial encoding latency[1,10,12,13].

Recently, the emerging field of synthetic dimension has introduced a transformative approach to constructing high-dimensional spaces by harnessing internal photonic degrees of freedom, such as frequency[14-17], time[18,19], mode[20], and orbital angular momentum[21]. Pioneering works have implemented convolutional operations in space combined with synthetic dimensions[22,23], thereby enabling parallel processing that surpasses time-domain limitations. Nonetheless, these demonstrations typically depend on resonant structures, which inherently restrict operational bandwidth and practical scalability. A fundamental question remains whether convolution can find natural embodiment in the dynamics of optical systems[24], which would be crucial for overcoming the constraints in system architecture and functionality. Yet, despite the long-standing pursuit of high-speed, low-complexity photonic computing architectures, this profound connection has remained unexplored.

In this work, we overcome this challenge by leveraging the universal principle that wave dynamics in translation-symmetric lattice systems intrinsically performs convolution. Inspired by this insight, we develop a photonic convolution architecture of exceptional simplicity through wave dynamics evolution in programmable synthetic frequency lattices. With a single electro-optic modulator and unshackled from the serial bottlenecks of time-domain approaches, we demonstrate high-throughput and multifunctional convolutional computation for diverse image tasks operating at a rate of 13.5 TOPS, showing great promise for realizing unconventional processing by engineering the synthetic lattice. More profoundly, the complex-valued convolution kernel inherent in the wave equation allows us to simulate not only classically irreversible diffusion process but also time-reversible, unitary quantum dynamics. Capitalizing on the physics-based reversibility and undetectable phase information, we develop a novel convolution-driven optical encryption strategy.

**Results**

**Universal principle for convolution from wave dynamics**. Consider a generic $n$-dimensional lattice with nearest-neighbor coupling, a paradigm encompassing a wide range of physical systems from photonic crystals to ultracold atoms in optical lattices. Within the tight-binding approximation, the wave packet evolution dynamics through the lattice is governed by the coupling-mode equation[14]

$$i\frac{\partial a(\boldsymbol{r},t)}{\partial t} = -\sum_{l=1}^{n} C_l \left[ a(\boldsymbol{r}+\boldsymbol{\Lambda}_l, t) + a(\boldsymbol{r}-\boldsymbol{\Lambda}_l, t) \right], \tag{1}$$

where $a(\boldsymbol{r},t)$ is the wave amplitude at lattice site $\boldsymbol{r}$. $\boldsymbol{\Lambda}_l$ is the unit vector in the $l$ direction, and $C_l$ is the corresponding coupling strength. To solve Eq. (1), we transition to the quasi-momentum ("$\boldsymbol{k}$") space via Fourier transform, which yields the evolution equation for Bloch states,

$$i\frac{\partial a(\boldsymbol{k},t)}{\partial t} = E(\boldsymbol{k})a(\boldsymbol{k},t), \tag{2}$$

where $E(\boldsymbol{k}) = -\sum C_l \cos(\boldsymbol{k}_l \boldsymbol{\Lambda}_l)$ denotes the band dispersion relation. This ordinary differential equation readily integrates to give the Bloch state solution at time $t$. By applying the inverse Fourier transform to real space and invoking the convolution theorem, we arrive at the formal solution for the wave function,

$$a(\boldsymbol{r},t) = a(\boldsymbol{r},0) * F^{-1}\left\{ \exp\left[-iE(\boldsymbol{k})\right] \right\}, \tag{3}$$

where $a(\boldsymbol{r},0)$ indicates the initial state, and $F^{-1}$ denotes the inverse Fourier transform.

Equation (3) reveals a core physical correspondence on the intrinsic connection between convolution computation and lattice wave dynamics. It suggests that the temporal evolution of a wave in a translation-invariant lattice is physically equivalent to a convolution between the initial state and a kernel $K$ dictated solely by the dispersion relation, as shown in the conceptual schematic of Fig. 1. This convolution formalism is fundamentally important, as it originates directly from the intrinsic translational symmetry of the lattice Hamiltonian and applies for arbitrary lattices (Figs. 1b-1d). Physically, the kernel is inherently complex-valued, a direct consequence of the unitary nature of wave propagation. This allows the system to perform not only real-valued but also complex-valued convolutions, fully leveraging the wave nature of the physical system. Most importantly, this universal mechanism is broadly applicable for diverse physical platforms such as optical waveguide arrays[12,25], synthetic temporal and frequency lattices[22], cold-atom arrays[26], and so on, establishing a unified framework where convolution is harnessed as a natural property of wave transport in translation-invariant systems.

**Experimental realization in synthetic frequency dimension.** The universal principle established in Section 1 transforms convolution implementation into the engineering of a lattice with a desired dispersion relation. In particular, the synthetic frequency dimension provides an ideal platform for optical computation and quantum simulation[6,27-29], where synthetic lattices are constructed by coupling equally spaced frequency modes. The coupling can be precisely controlled via electro-optic modulation[17], realizing a tight-binding model and enabling access to the convolution kernel through the underlying lattice dynamics. For a one-dimensional (1D) frequency lattice constructed by sinusoidal phase modulation $V = V_0\cos(\Omega t)$, the dynamics under tight-binding approximation is described by

$$a(p,t) = a(p,0) * F^{-1}\left\{\exp[im\cos(\Omega k)]\right\}, \qquad (4)$$

where $p$ indexes the lattice sites, and $\Omega$ is the modulation frequency, equal to the lattice period. Crucially, the convolution kernel emerges as $k_p = i^p J_p(m)$, derived from the dispersion relation $E(k) = -m\cos(\Omega k)$, where $J_p(m)$ is the $p$-order Bessel function, $m = \pi V_0/V_\pi$ is the modulation depth, and $V_\pi$ is the half-wave voltage. This offers inherent programmability since the kernel is uniquely determined by the dispersion relation, which can be governed through the modulation signal in synthetic frequency lattices[30].

Strikingly, the principle enables to develop a convolution architecture of exceptional simplicity. As illustrated in Fig. 2a, the entire computation is accomplished with a single electro-optic modulator. Input data is encoded as the initial amplitude distribution across a frequency comb. The natural evolution of light under this programmed modulation inherently performs convolution in the synthetic frequency space. Entirely operating in the frequency domain, this approach enables full parallelism and circumvents the serial bottleneck of time-domain schemes (Fig. 2b). Experimentally, a 1D data sequence is encoded onto an incoherent frequency comb ($\Omega/2\pi = 25$ GHz, Fig. 3a), realized by programmable filtering of a broadband amplified spontaneous emission (ASE) source with a waveshaper[3]. The utilization of an incoherent frequency comb relaxes the requirement for complex coherent combs and supports massive encoding bandwidth. Then, convolution is executed using an electro-optic Mach–Zehnder amplitude modulator (AM). Under small-signal modulation, a compact 3-element intensity kernel can be obtained as the fourth-order Bessel function approaches zero,

$$K_{AM} = [J_{-2}^2(m), J_0^2(m), J_2^2(m)]. \tag{5}$$

By tuning the modulation depth $m$ to match a Gaussian kernel [1, 2, 1], the experimental output in Fig. 3b demonstrates a clear Gaussian blurring of the input signal. The versatility is further demonstrated by performing image processing on a 30×30 pixel image (digit "7"), as presented in Fig. 3c. By programming $m$ to achieve convolution kernels of [0, 1, 0] ($m = 0$), [1, 2, 1] ($m = 0.541\pi$), and [1, 1, 1] ($m = 0.586\pi$), multiple functions can be respectively produced in the horizonal and vertical directions, including identity mapping, blurring, and mean filtering. In addition, edge detection (kernel [-1, 2, -1]) is realized via linear post-processing, which clearly highlights the digit features.

To extend to two-dimensional (2D) processing without escalating hardware complexity, an innovative strategy is demonstrated with the same simple architecture. Generally, 2D convolution can be performed using a synthetic 2D lattice (Fig. 1c), yet a high-speed modulation (approximately 12.5×30 = 375 GHz, for a 30×30 pixel image) is required for constructing long-range coupling[31]. Exploiting kernel separability instead, 2D convolution can be efficiently decomposed into sequential 1D evolutions along the horizontal and vertical directions (Fig. 4a)[32],

$$f * K_{2D} = f * K_X * K_Y. \tag{6}$$

This method is validated by producing various standard kernels for processing 2D images. As shown in Fig. 4b, 2D Gaussian blurring of the digit "7" image is realized by setting $K_{X1} = [1, 2, 1]$, $K_{Y1} = [1, 2, 1]^T$, and mean filtering of the digit "8" image is realized by setting $K_{X2} = [1, 1, 1]$, $K_{Y2} = [1, 1, 1]^T$, respectively. Besides, sharpening of the flower image and edge detection on the Tai Chi image can be also demonstrated via post processing. Comparisons showcase that all the results obtained via 2D frequency convolution agree well with those calculated by a computer (Supplementary Note 1). Furthermore, we propose that the general principle relating convolution and lattice wave dynamics is beneficial for realizing unconventional processing tasks, by engineering the synthetic lattice with advanced modulation schemes (Supplementary Note 2)[17,33]. As demonstrated in Fig. 4c, a flower image with triangular-pixel arrangement can be directly processed through convolution within a synthetic triangular lattice (Fig. 1d), implying extensive programmability and scalability.

Specifically, the processing rate of this convolutional architecture can be determined by

$$S = 2N_{\text{ker}}N_{data}\frac{1}{\Delta t} = 2N_{\text{ker}}W_f,  \qquad (7)$$

where $N_{\text{ker}}$ is the size of convolution kernel, $N_{data}$ is the encoding comb lines, and $\Delta t = 2\pi/\Omega$ is the processing time. The spectral width for signal coding can be calculated as $W_f = N_{data}\Omega/2\pi$. Notably, the computational speed is independent on the modulation frequency, but scales linearly with the spectral width. In our experiment, $N_{\text{ker}} = 3$, and the encoding width is $W_f = 2.25$ THz, yielding a processing rate of 13.5 TOPS. The minimal architecture, based on a single electro-optic modulator, is fully compatible with emerging thin-film lithium niobate platforms[30,34]. With large-scale on-chip integration, the processing rate could be further enhanced by orders of magnitude[34,35].

**Convolution-driven diffusion simulation and optical encryption.** The connection to wave dynamics not only enables computation but also opens a pathway for simulating complex physical processes. From the perspective of wave dynamics, the temporal evolution (convolution) governed by Eq. (3) represents a generalized diffusion process within the synthetic lattice. To simulate classical particle diffusion described by $u(x, y, t) = u(x, y, 0)*F^{-1}[\exp(-Dk_x^2 t)]*F^{-1}[\exp(-Dk_y^2 t)]$ where $D$ is the diffusion coefficient and $t$ is the diffusion time, an incoherent initial state is processed by performing separable frequency convolutions in $X$ and $Y$ directions with an AM (Fig. 5a). The resulting kernel, $K = F\{\cos[m\cos(\Omega t)]\}$, corresponds to an inherently irreversible diffusion as the inverse kernel $F\{1/\cos[m\cos(\Omega t)]\}$ diverges. By controlling the modulation depth $m$ to emulate varying diffusion constants $D$, the diffusion effects of a point source and a multiparticle distribution are accurately simulated at stages $Dt = 0$, 0.3 and 0.6, as shown in (i-iii) and (iv-vi) of Fig. 5b respectively.

More significantly, the quantum-mechanical treatment of the wave dynamics reveals a deeper capability. The system evolves under a Schrödinger-type equation with a complex-valued convolution kernel. The time-reversal symmetry implies that the underlying quantum process is fundamentally reversible. However, the use of an incoherent initial state, which lacks a detectable phase, renders the actual evolution a classical-like diffusion process. To experimentally access to the quantum nature, two cascaded phase modulators (PMs) are used to simulate the diffusion and recovery (Fig. 5c). The first PM performs the convolution with kernel $K = F\{\exp[im\cos(\Omega t)]\}$, corresponding to a unitary evolution operator $U$. While the second PM applies the exact inverse $K^{-1} = F\{\exp[-im\cos(\Omega t)]\}$ (Hermitian conjugate $U^\dagger$), thereby implementing the time-reversal operation $U^\dagger U$. Fig. 5d shows that both a point source and a structured pattern (triangle) are nearly perfectly recovered after the full reversible process, demonstrating the successful simulation of time-reversible unitary quantum dynamics within our architecture.

Leveraging this physics-based reversibility and the undetectable phase information of incoherent light, a unique paradigm is developed for optical encryption. As schematized in Fig. 6a, plaintext data interleaved with random noise ("salt") is encoded onto a frequency lattice. The first PM diffuses the data and noise together via the complex kernel, physically scrambling the information into ciphertext. The absence of a measurable phase in the incoherent frequency comb makes it impossible for an interceptor to invert the process mathematically, even with full knowledge of the system architecture. While for decryption, a precise inverse kernel is required to serve as the physical cipher. Applied via the second PM, the diffusion process is reversed and the original information can be recovered with high fidelity. Figs. 6b-6d present the experimental demonstrations for a digital signal encoding "HUST" (binary 1001000 1010101 1010011 1010100). The ciphertext (Fig. 6c) shows complete distortion against the original data (Fig. 6b), while the decrypted signal (Fig. 6d) reproduces the original with high accuracy. Similar optical encryption and decryption are also performed with analog signals (Supplementary Note 3), demonstrating high-security encryption and precise recovery. It is important to emphasize that although a simple sinusoidal modulation is used for proof of principle, the security can be further enhanced with complex modulation signals. Overall, this work establishes a novel diffusion-based encryption strategy, offering exceptional resilience and a novel approach for secure information processing.

## Discussion

We have revealed a fundamental correspondence between wave dynamics in physical lattices and convolutional computation, and validated its implementation with a simple photonic architecture. While photonic convolutions have been previously realized using various hardware platforms, our work is distinguished by the foundation in a universal physical principle and the minimal complexity, yielding several key advances. First, the general principle motivates to realize convolution with arbitrary synthetic lattices, showcasing extensive programmability and scalability for unconventional processing tasks. Second, this work harnesses intrinsic parallelism overcoming the serial time-domain limitations, and the use of an incoherent frequency comb eliminates the need for complex coherent frequency comb sources while enabling massive bandwidth utilization that directly translates to high computational throughput[3]. Most significantly, the complex-valued kernels preserve the quantum nature even with classical incoherent light, which permits simulation of unitary quantum dynamics and inspire a novel encryption strategy based on physical dynamics rather than mathematical algorithms. Moreover, this universal principle is platform-agnostic and can be transferred to other systems such as metamaterials[36], cold-atom arrays[26], and superconducting circuits[37], offering a unified framework for optical computation across physics.

In conclusion, we have established that wave dynamics in translation-symmetric lattices physically constitutes a convolution operation. This universal principle enables the development of a convolution architecture of minimal complexity in synthetic dimension, achieving a high-throughput and multifunctional photonic processor operating at a rate of 13.5 TOPS for image processing tasks. The single-modulator design contrasts favorably with previous complex systems, highlighting excellent integration potential and energy efficiency. Moreover, utilizing the kernel's complex nature, we realize the simulation of classically irreversible diffusion and reversible unitary quantum dynamics, and demonstrate a novel encryption strategy based on photonic convolution. This work establishes a unified framework for photonic computing, opening avenues toward scalable multifunctional photonic processors with high integration potential for optical neural network, quantum simulation, and secure communication.

**References**

[1] X. Y. Xu, M. X. Tan, B. Corcoran, J. Y. Wu, A. Boes, T. G. Nguyen, S. T. Chu, B. E. Little, D. G. Hicks, R. Morandotti, A. Mitchell, and D. J. Moss, Nature **589**, 44 (2021).

[2] J. Feldmann, N. Youngblood, M. Karpov, H. Gehring, X. Li, M. Stappers, M. Le Gallo, X. Fu, A. Lukashchuk, A. S. Raja, J. Liu, C. D. Wright, A. Sebastian, T. J. Kippenberg, W. H. P. Pernice, and H. Bhaskaran, Nature **589**, 52 (2021).

[3] B. Dong, F. Brückerhoff-Plückelmann, L. Meyer, J. Dijkstra, I. Bente, D. Wendland, A. Varri, S. Aggarwal, N. Farmakidis, M. Wang, G. Yang, J. S. Lee, Y. He,

E. Gooskens, D.-L. Kwong, P. Bienstman, W. H. P. Pernice, and H. Bhaskaran, Nature **632**, 55 (2024).

[4] F. Stelzer, A. Röhm, R. Vicente, I. Fischer, and S. Yanchuk, Nat. Commun. **12**, 5164 (2021).

[5] X. Meng, N. Shi, G. Zhang, J. Li, Y. Jin, S. Sun, Y. Shen, W. Li, N. Zhu, and M. Li, Light: Sci. Appl. **14**, 27 (2025).

[6] B. Bai, Q. Yang, H. Shu, L. Chang, F. Yang, B. Shen, Z. Tao, J. Wang, S. Xu, W. Xie, W. Zou, W. Hu, J. E. Bowers, and X. Wang, Nat. Commun. **14**, 66 (2023).

[7] Y. Wang, K. Liao, K. Zhang, Z. Du, Z. Wang, B. Ni, T. Xu, S. Feng, Y. Yang, Q.-F. Yang, Q. Sun, X. Hu, and Q. Gong, eLight **5**, 20 (2025).

[8] Y. Shen, N. C. Harris, S. Skirlo, M. Prabhu, T. Baehr-Jones, M. Hochberg, X. Sun, S. Zhao, H. Larochelle, D. Englund, and M. Soljačić, Nat. Photonics **11**, 441 (2017).

[9] Z. Du, K. Liao, T. Dai, Y. Wang, J. Gao, H. Huang, H. Qi, Y. Li, X. Wang, X. Su, X. Wang, Y. Yang, C. Lu, X. Hu, and Q. Gong, Sci. Adv. **10**, eadm7569 (2024).

[10] Y. Bai, Y. Xu, S. Chen, X. Zhu, S. Wang, S. Huang, Y. Song, Y. Zheng, Z. Liu, S. Tan, R. Morandotti, S. T. Chu, B. E. Little, D. J. Moss, X. Xu, and K. Xu, Nat. Commun. **16**, 292 (2025).

[11] W. T. Gu, X. Y. Gao, W. C. Dong, Y. L. Wang, H. L. Zhou, J. Xu, and X. L. Zhang, Optica **11**, 64 (2024).

[12] X. Meng, G. Zhang, N. Shi, G. Li, J. Azaña, J. Capmany, J. Yao, Y. Shen, W. Li, N. Zhu, and M. Li, Nat. Commun. **14**, 3000 (2023).

[13] X. Zhang, Z. Sun, Y. Zhang, J. Shen, Y. Chen, M. Sun, C. Shu, C. Zeng, Y. Jiang, Y. Tian, J. Xia, and Y. Su, Laser Photon. Rev. **19**, 2401583 (2025).

[14] L. Q. Yuan, Q. Lin, M. Xiao, and S. H. Fan, Optica **5**, 1396 (2018).

[15] A. Dutt, Q. Lin, L. Q. Yuan, M. Minkov, M. Xiao, and S. H. Fan, Science **367**, 59 (2020).

[16] D. Yu, G. Li, L. Wang, D. Leykam, L. Yuan, and X. Chen, Phys. Rev. Lett. **130**, 143801 (2023).

[17] A. Senanian, L. G. Wright, P. F. Wade, H. K. Doyle, and P. L. McMahon, Nat. Phys. **19**, 1333 (2023).

[18] A. Regensburger, C. Bersch, M. A. Miri, G. Onishchukov, D. N. Christodoulides, and U. Peschel, Nature **488**, 167 (2012).

[19] S. Wang, C. Qin, W. Liu, B. Wang, F. Zhou, H. Ye, L. Zhao, J. Dong, X. Zhang, S. Longhi, and P. Lu, Nat. Commun. **13**, 7653 (2022).

[20] E. Lustig, S. Weimann, Y. Plotnik, Y. Lumer, M. A. Bandres, A. Szameit, and M. Segev, Nature **567**, 356 (2019).

[21] M. Yang, H. Q. Zhang, Z. H. Liu, Z. W. Zhou, X. X. Zhou, J. S. Xu, Y. J. Han, C. F. Li, and G. C. Guo, Sci. Adv. **9**, eabp8943 (2023).

[22] L. L. Fan, K. Wang, H. M. Wang, A. Dutt, and S. H. Fan, Sci. Adv. **9**, eadi4956 (2023).

[23] H. Zheng, Q. Liu, Y. Zhou, I. I. Kravchenko, Y. Huo, and J. Valentine, Sci. Adv. **8**, eabo6410 (2022).

[24] M. Favoni, A. Ipp, D. I. Müeller, and D. Schuh, Phys. Rev. Lett. **128**, 032003 (2022).

[25] J. Cheng, C. Li, J. Dai, Y. Chu, X. Niu, X. Dong, and J.-J. He, Laser Photon. Rev. **18**, 2301221 (2024).

[26] M. Reinschmidt, J. Fortágh, A. Günther, and V. V. Volchkov, Nat. Commun. **15**, 8532 (2024).

[27] D. Cheng, K. Wang, C. Roques-Carmes, E. Lustig, O. Y. Long, H. Wang, and S. Fan, Nature **637**, 52 (2025).

[28] H. Zhao, B. Li, H. Li, and M. Li, Nat. Commun. **13**, 5426 (2022).

[29] F. Pellerin, R. Houvenaghel, W. A. Coish, I. Carusotto, and P. St-Jean, Phys. Rev. Lett. **132**, 183802 (2024).

[30] W. Liu, X. Su, C. Li, C. Zeng, B. Wang, Y. Wang, Y. Ding, C. Qin, J. Xia, and P. Lu, Phys. Rev. Lett. **134**, 143801 (2025).

[31] L. Fan, Z. Zhao, K. Wang, A. Dutt, J. Wang, S. Buddhiraju, C. C. Wojcik, and S. Fan, Phys. Rev. Appl. **18**, 034088 (2022).

[32] J. S. Lim, *Two-Dimensional Signal and Image Processing* (Prentice Hall, Englewood Cliffs, 1990).

[33] K. Wang, A. Dutt, K. Y. Yang, C. C. Wojcik, J. Vuckovic, and S. H. Fan, Science **371**, 1240 (2021).

[34] H. Feng, T. Ge, X. Guo, B. Wang, Y. Zhang, Z. Chen, S. Zhu, K. Zhang, W. Sun, C. Huang, Y. Yuan, and C. Wang, Nature **627**, 80 (2024).

[35] C. Wang, M. Zhang, X. Chen, M. Bertrand, A. Shams-Ansari, S. Chandrasekhar, P. Winzer, and M. Loncar, Nature **562**, 101 (2018).

[36] F. Zangeneh-Nejad, D. L. Sounas, A. Alù, and R. Fleury, Nat. Rev. Mat. **6**, 207 (2021).

[37] I. Cong, S. Choi, and M. D. Lukin, Nat. Phys. **15**, 1273 (2019).

## Methods

**Experimental details.** A broadband amplified spontaneous emission (ASE) source (Shanghai B&A Technology) is used as the light source. A programmable optical filter (WSS-2000, Santec) shapes the source into an incoherent frequency comb and assigns corresponding intensity of the loaded data onto the comb lines. For image processing tasks and simulation of diffusion processes, the encoding intervals are set as 25 GHz and 8 GHz, respectively. Commercial electro-optic phase modulator (MPZ-LN-20, iXblue) and Mach–Zehnder amplitude modulator (MX-LN-10, Exail) are used to perform photonic convolution in synthetic frequency dimension, driven by radio-frequency (RF) signals from a microwave source (SMF 100A, Rohde&Schwarz). The phase difference between the modulated signals is controlled by a phase shifter (model 981, Weinschel). The output spectra are measured using an optical spectrometer (AQ6370B, YOKGAWA). To monitor the convolution kernel, a frequency line far from the data region is encoded on the comb as a reference.

## Data availability

The data that support the findings of this study are available from the corresponding authors upon request.

## Acknowledgements

This work was supported by the National Key Research and Development Program of China (Grant No. 2023YFA1406800) and National Natural Science Foundation of China (No. 12374305, No. 62375097, No. 12021004).

## Author contributions

W. L., P. L. and B. W. developed the concept and guided the project. X. S., W. L., R. C. and H. Z. performed the experiment measurements, data analysis and the numerical simulations. All the authors contributed to discussion and writing of the manuscript.

## Competing interests

The authors declare no competing interests.

## Additional information

**Supplementary information** The online version contains supplementary material available at https://doi.org/xxxxxx.

**Figures and captions**

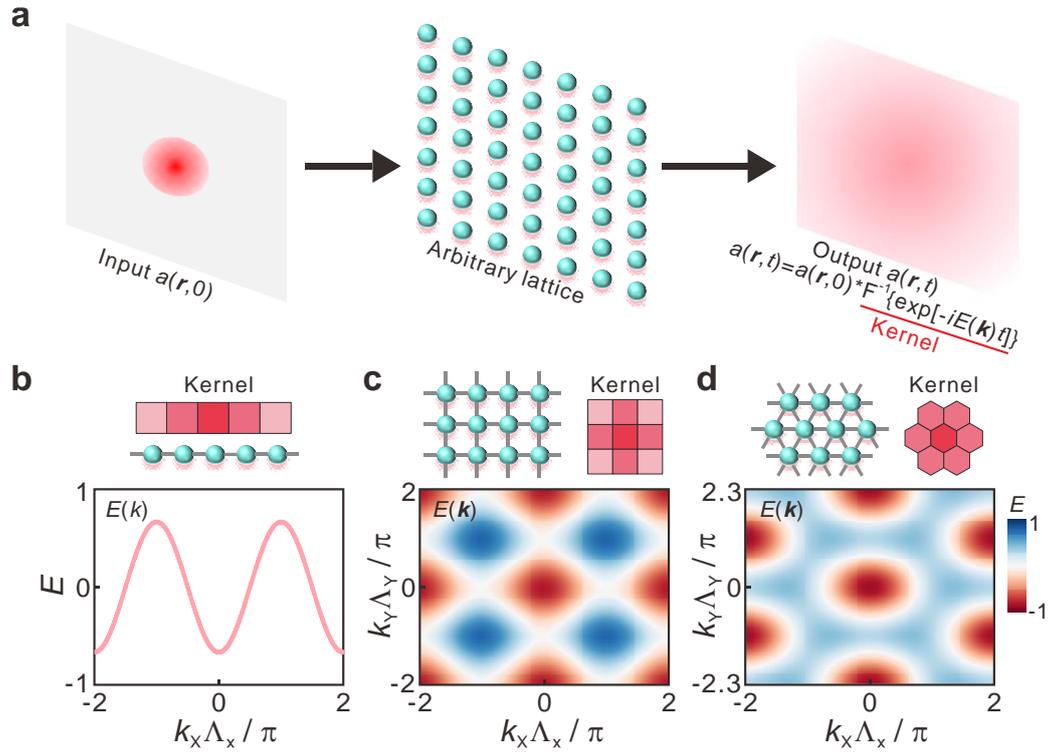

**Fig. 1 Concept of convolution from wave dynamics in lattices. a**, Illustration for performing convolution through wave dynamics evolution in physical lattices. The temporal evolution of a wave in a translation-invariant lattice is physically equivalent to a convolution between the initial state and a kernel, $K = F^{-1}\{\exp[-iE(\mathbf{k})t]\}$, dictated by the dispersion relation $E(\mathbf{k})$. **b-d**, Structures, energy bands and corresponding convolution kernels for typical one-dimensional (1D), two-dimensional (2D) square and triangular lattices. In principle, the convolution formalism applies for arbitrary lattices as it originates from the intrinsic translational symmetry of the lattice Hamiltonian.

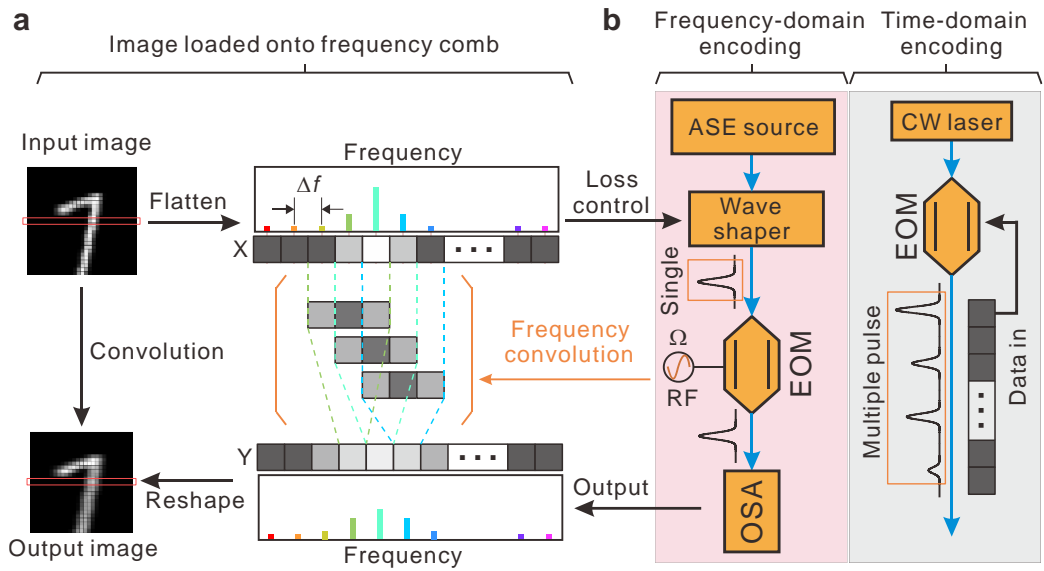

**Fig. 2 Experimental implementation of convolution in synthetic frequency dimension. a**, Structural schematic for performing convolution with a single electro-optic modulator. Input data is flattened into a one-dimensional sequence and encoded on an incoherent frequency comb using wave shaper, which is then introduced into an electro-optic modulator (EOM) to perform convolution in the synthetic frequency space. The output spectra are measured with an optical spectrum analyzer (OSA). **b**, Comparison of parallel frequency encoding and serial time-domain scheme. ASE source, amplified spontaneous emission source; CW laser, continuous-wave laser; RF, radio frequency.

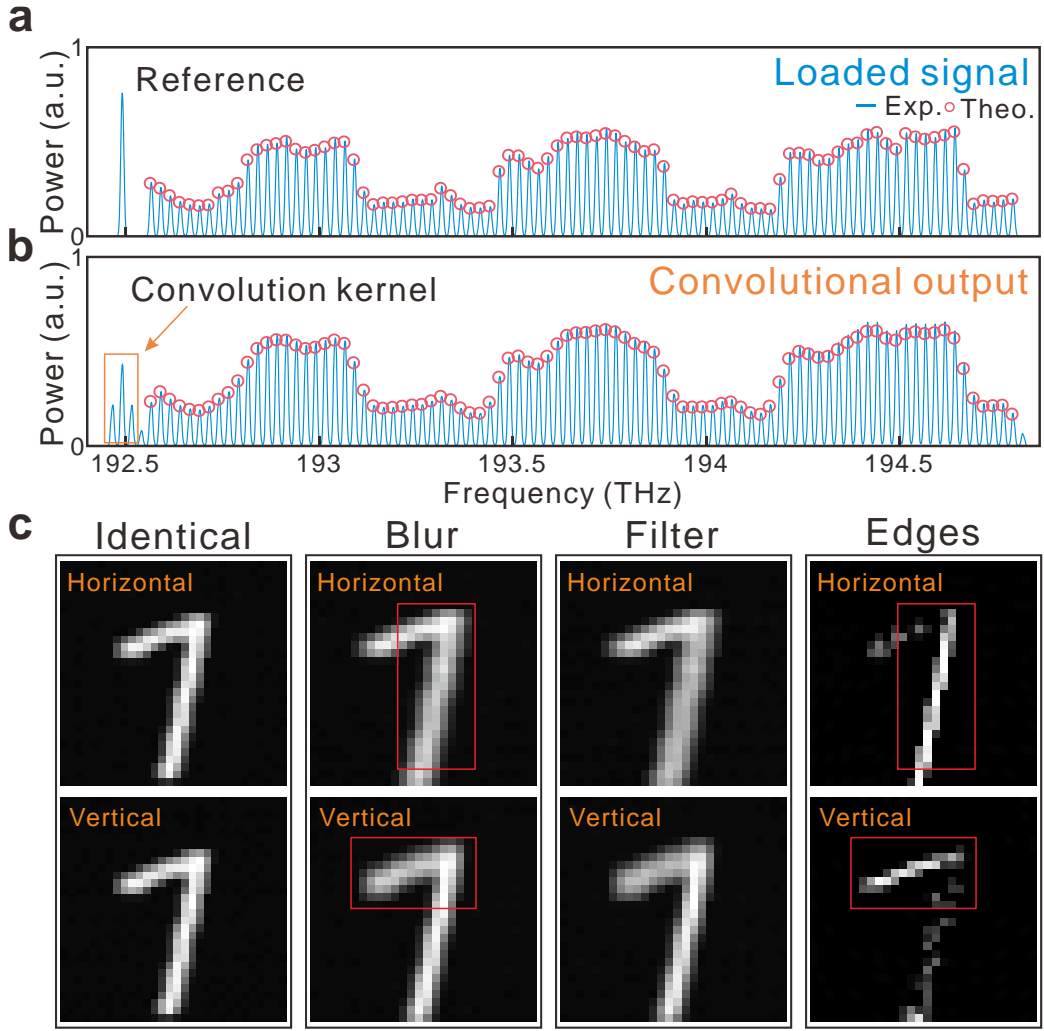

**Fig. 3 One-dimensional (1D) convolution in synthetic frequency dimension. a**, Loaded signal of a flattened 1D data sequence encoded on an incoherent frequency comb. A frequency line far from the data region is encoded on the comb to act as a reference. **b**, Corresponding convolutional output of the data sequence in (**a**), with a kernel [1, 2, 1]. The convolution kernel shown on the left side originates from the evolution of the single frequency line. **c**, Image processing results for identity mapping ($K_1 = [0, 1, 0]$), blurring ($K_2 = [1, 2, 1]$), mean filtering ($K_3 = [1, 1, 1]$), and edge detecting ($K_4 = [-1, 2, -1]$), in the horizontal and vertical directions respectively. The diverse convolution kernels are generated by programming modulation depth $m$ ($m = 0$, $0.541\pi$, and $0.586\pi$ respectively) to manipulate dispersion relation of the synthetic frequency lattice. Edge detection is realized via linear post-processing, $K_4 = 3K_1 - K_3$.

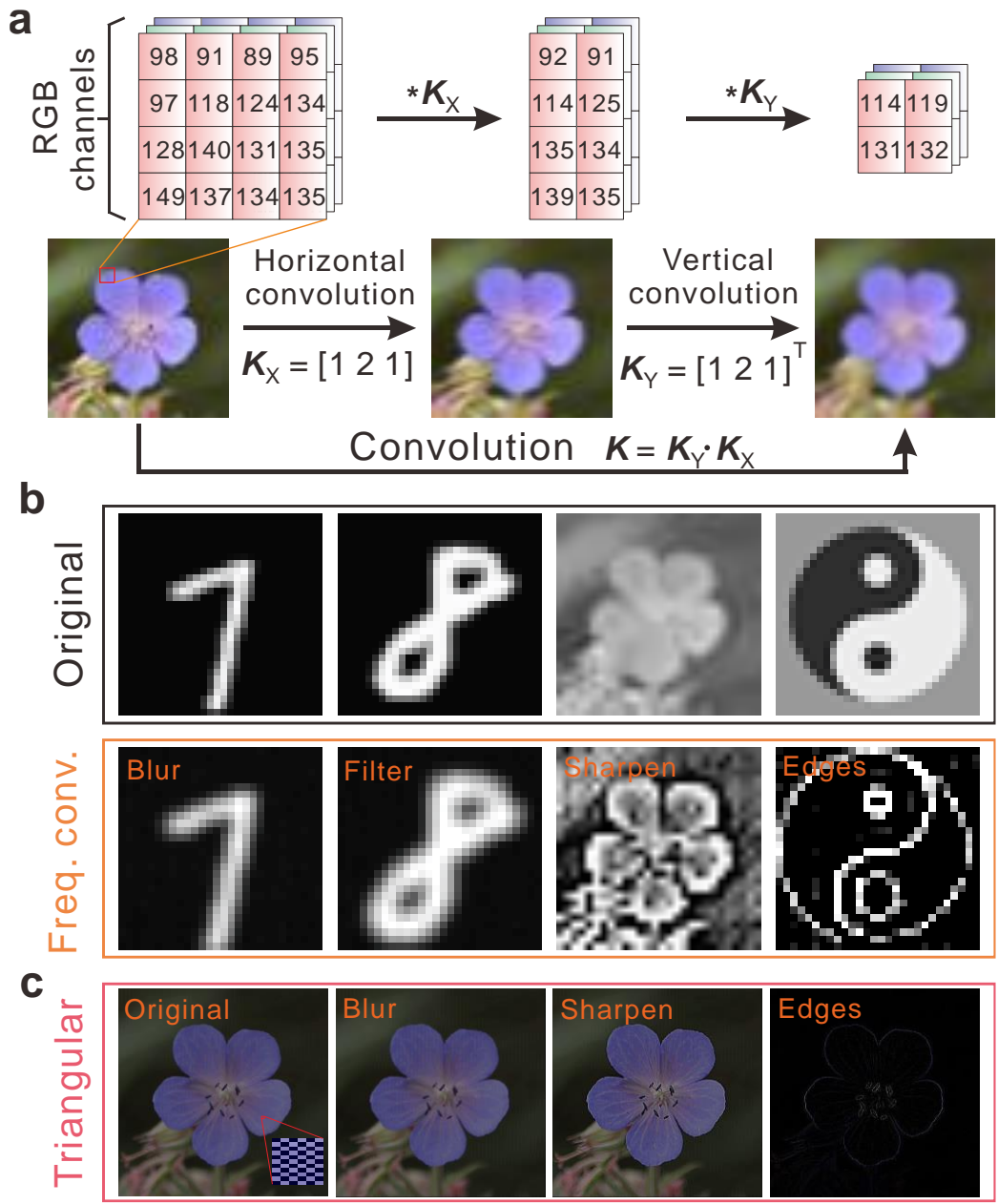

**Fig. 4 Two-dimensional (2D) convolution for image processing in synthetic frequency dimension.**
**a**, Schematic illustration for 2D convolution exploiting kernel separability, $f*K_{2D} = f*K_X*K_Y$. **b**, The original images, and corresponding processing results for blurring, mean filtering, sharpening and edge detecting using 2D convolution, which is realized by performing horizontal and vertical convolutions in sequence. **c**, Calculated processing results for a flower image with triangular-pixel arrangement (enlarged illustration in the inset) through convolution within a synthetic triangular lattice. This proposal indicates that the extensive programmability and scalability of convolution in synthetic lattice is beneficial for directly processing unconventional tasks, without relying on complex algorithms.

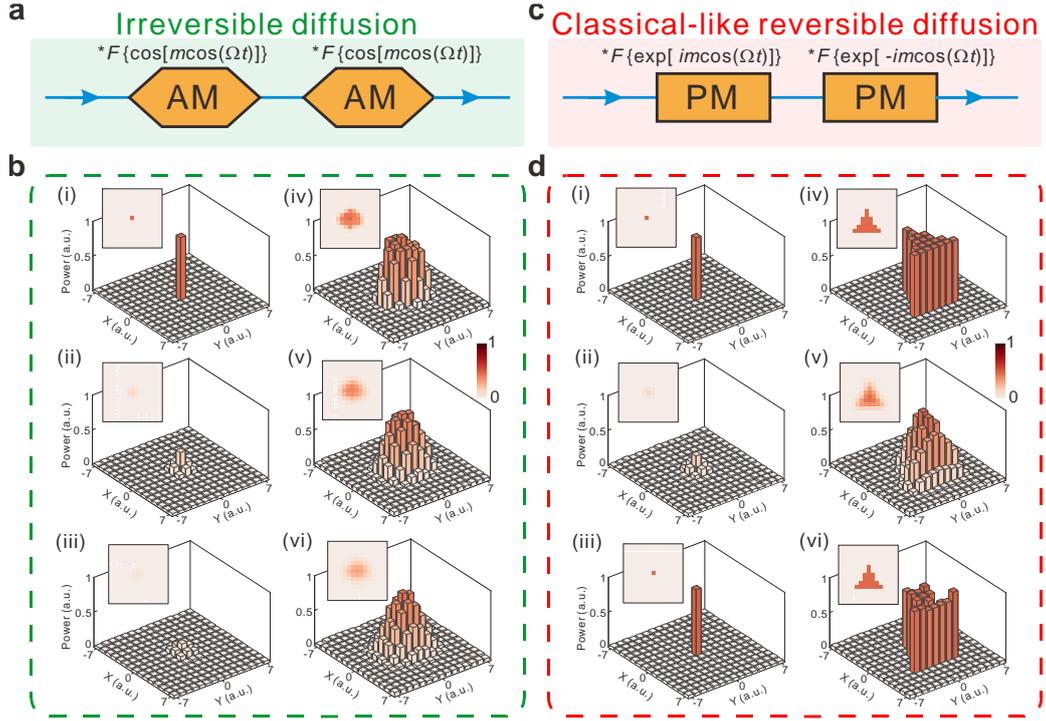

**Fig. 5 Convolution-driven photonic simulation of physical diffusion processes. a**, Schematic illustration of architecture for simulating classical irreversible diffusion. The electro-optic amplitude modulator (AM) performs convolution with kernel $K = F\{\cos[m\cos(\Omega t)]\}$, while the inverse kernel $F\{1/\cos[m\cos(\Omega t)]\}$ diverges, corresponding to an inherently irreversible process. **b**, Experimental results for diffusion simulations of a point source (i-iii) and a multiparticle distribution (iv-vi), at $Dt = 0$, 0.3 and 0.6, respectively. **c**, Schematic illustration of architecture for simulating classical-like reversible unitary quantum dynamics. The first electro-optic phase modulator (PM) performs convolution with kernel $K = F\{\exp[im\cos(\Omega t)]\}$, corresponding to a unitary evolution operator $U$. The second PM applies inverse $K^{-1} = F\{\exp[-im\cos(\Omega t)]\}$, corresponding to the Hermitian conjugate $U^\dagger$. **d**, Experimental results for diffusion and recovery simulation of a point source (i-iii) and a triangle (iv-vi), respectively. Insets present corresponding theoretical calculated results.

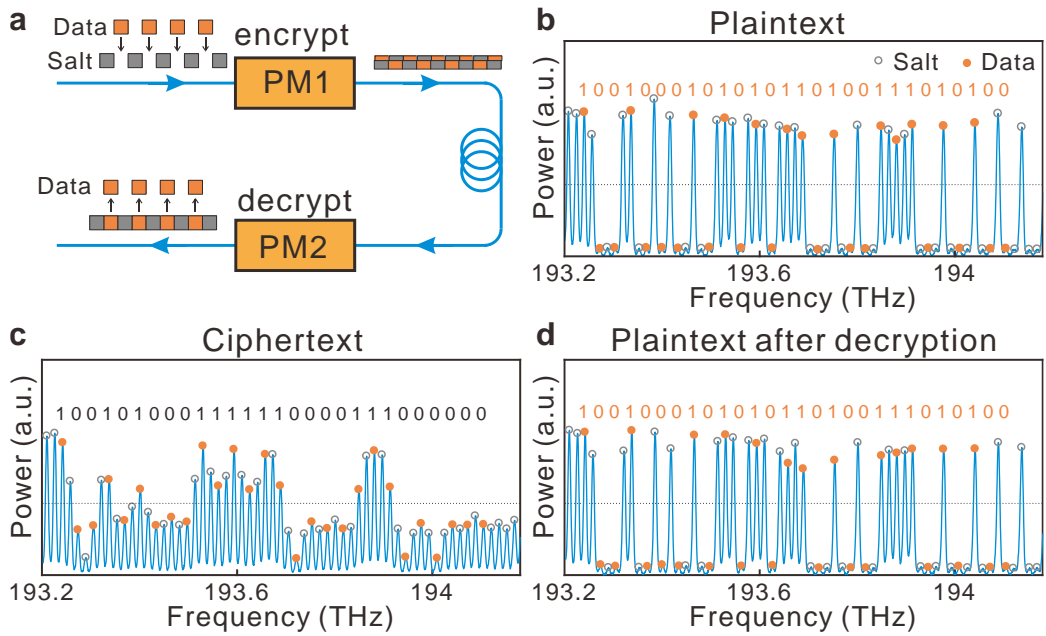

**Fig. 6 Convolution-driven optical encryption. a**, Schematic illustration for optical encryption and decryption utilizing convolution in synthetic frequency dimension. Plaintext data interleaved with random noise ("salt") is encoded onto a frequency lattice. The first PM diffuses the data and noise together to perform encryption. The undetectable phase information in the incoherent frequency comb makes it impossible to invert the process mathematically, thus providing a high security. The second PM performs decryption by recovering the original information with an inverse kernel. **b**, Experimental plaintext data interleaved with random noise ("salt"). **c**, Ciphertext after physically scrambling through photonic convolution. **d**, Plaintext after decryption with an inverse convolution.